\documentclass{article}
\usepackage[a4paper,margin=2.2cm]{geometry}
\usepackage{rotating}
\usepackage{epsf}

\title{Some static spherically symmetric interior solutions of $f(R)$ gravity}
\author{A. Shojai and F. Shojai\\
Department of Physics, University of Tehran, Tehran, Iran.} 
\date{}
\begin{document}
\maketitle
\begin{abstract}
We have found some new exact static spherically symmetric interior solutions of metric $f(R)$ gravitational theories describing the equilibrium configuration of a star. Then  the solution is matched to the exterior solution and thus gives a complete description of a star in $R^{1-n/2}$ theory.
\end{abstract}
\section{Introduction}
One of the important problems in the gravitational theories is finding solutions under the assumption of spatial isotropy and time independence, i.e. static spherically symmetric solutions. Then the results can be used to treat the solar system tests and also for computing the pressure, density and gravitational field within a spherically symmetric static star. Regarding the standard general relativity, there is a large number of exact solutions of Einstein's field equations for perfect spherically symmetric static fluids in the literatures. For a complete list see \cite{lake1}. From these solutions which are in closed algebraic form, only eleven solutions satisfy physical criteria mentioned in \cite{lake1} and therefore are physically acceptable. (nine solutions in \cite{lake1}, one in \cite{lake2} and one in \cite{kiess}).

A particular class of gravitational theories which has been received great interest recently, is $f(R)$ modified theories of gravity \cite{fr}. In \cite{vil6} spherically symmetric vacuum solutions of $f(R)$ theories are studied. In particular it is shown that the modified gravitational theories have the general static blackhole solutions with non-vanishing cosmological constant, Schwarzschild--deSitter solution\cite{vil6, one, En}. Considering scalar--tensor theories of gravity in the Einstein frame, a reconstruction method is introduced in \cite{on} to generate some spherically symmetric vacuum space--times. Since $f(R)$ gravity is dynamically equivalent to Brans-Dicke theory with zero coupling constant, this method is applicable to $f(R)$ gravity in the Einstein frame. Also in the recent years some exact solutions for static spherically symmetric perfect fluid in $f(R)$ theory of gravity are presented \cite{vilst, kain, cap}. Since the field equations of $f(R)$ theory are differential equations of fourth order, more integration constants need to be determined in order to get a unique solution. This shows that in general, the exterior space-time geometry is not described only by the mass of the star, and thus the validity of Birkhoff theorem is lost\cite{clif1,clif2}. Other details of mass distribution, i.e. the gravitational multipole moments, are needed to obtain a unique solution for both interior and exterior regions of a star. As a result  the exterior solutions of $f(R)$ theory are not necessarily asymptotically flat.

Another issue related to the star evolution is the singularity problem in $f(R)$ gravity which has been first introduced in \cite{barr} and then disscused for stellar configurations in \cite{Kob, Bam} which maybe treated by adding higher powers of $R$ \cite{Bamb}. The classification of 
future singularities according to their properties into four classes is studied in \cite{3}. These singularities occur in scalar or fluid models of dark energy and in $f(R)$ or $f(G)$ (G denotes the Gauss-
Bonnet invariant) \cite{4} or other models leading to an  accelerating universe. The fact that all such singularities is not unique for $f(R)$ but occur in all known dark energy models is explained in \cite{Bamb, 5}.

Moreover the consistency condition of any $f(R)$ theory with the solar system tests is the occurrence of the chameleon mechanism \cite{Kh,Hu}. It is one way to avoid the Doglov--Kawaski instability \cite{Dog}. Also a positive value of $F(R)=df(R)/dR$ is needed to guarantee the positivity of the effective gravitational constant $G_{eff}=G/F(R)$ \cite{fr}. 

In \cite{clif1,clif2}, by considering the Lagrangian of the form of $R^{1+\delta}$, two exact 
spherically symmetric vacuum solutions are presented which do not have the asymptotic behavior of the Minkowski space-time. It seems that the non-existence of asymptotic flatness is one of the reasons to study weather $f(R)$ theory can explain the flat rotation curve of galaxies or not \cite{clif1, soboti, bohmer, Rahvar}. Moreover  $f(R)$ static spherically symmetric solution for stellar systems has been investigated in \cite{vil7}. The authors show that specifying the pressure and the density of matter do not uniquely determine the functional form of $f(R)$. As mentioned before, this is because of appearing higher order derivatives of quantities in the equations which require matching of higher order derivatives of variables at the surface of star. It is shown that assuming Schwartzschild--deSitter metric outside the mass distribution as a boundary condition, induces tight constraints on the quantities\cite{vil7}.

In this paper, we discuss the gravitational equations of $f(R)$ theory for the interior of a spherically symmetric matter distribution and derive a few new solutions in some closed algebraic form. For these solutions, the dependence of $f(R)$ function to the scalar curvature can be found. We divide these solutions into two groups. After analyzing each metric, we see that one of the obtained metrics is appropriate for describing a star. At the end, we consider this solution in more detail and match it to the corresponding exterior metric solution.  
\section{Spherically symmetric static solutions of $f(R)$ gravity}
The dynamics of $f(R)$ theory can be determined by the following action:
\begin{equation}
S=\int d^{4}x
\sqrt{-g}(f(R))+\kappa\mathcal{L}_{m})
\end{equation}
In the metric approach, the gravitational field equations can be obtained via the variation of action with respect to metric:
\begin{equation}
F(R) R_{\mu\nu}-
\frac{1}{2}f(R)g_{\mu\nu}-\nabla_{\mu}\nabla_{\nu}F(R)+g_{\mu\nu}\nabla^\alpha\nabla_\alpha{F(R)}=\kappa T_{\mu\nu}
\label{fr}
\end{equation}
where $T_{\mu\nu}$ is the energy-momentum tensor of matter and $F(R)=df/dR$. Considering a static and spherically symmetric distribution of matter, the line element has the following form:
\begin{equation}
ds^2=A(r)dt^2-B(r)dr^2-r^2d\Omega^2
\label{ds}
\end{equation}
and assuming the matter to be described by a perfect fluid, we have:
\begin{equation}
T^{\mu\nu}=(\rho+p)u^\mu u^\nu -p g^{\mu\nu}
\label{t}
\end{equation}
where $\rho(r)$ and $p(r)$ are the proper mass density and the isotropic pressure of the fluid in the instantaneous rest frame of the fluid and $u^\mu$ is its four velocity. Substituting (\ref{ds}) and (\ref{t}) in (\ref{fr}), one obtains the static spherically symmetric field equations of $f(R)$ theory with matter. Assuming a specific form of $f(R)$, these equations would be three fourth order nonlinear differential equations. Supplementing by an equation of state for matter:
\begin{equation}
p=p(\rho)
\label{state}
\end{equation}
one has a complete set of equations for four unknown quantities, $A$, $B$, $p$ and $\rho$. To carry out the calculation, it is desirable to re-express the equations in a somewhat different form \cite{vil6}. It can be shown that the expression:
\begin{equation}
K_{[\mu]}=\frac{FR_{\mu\mu}-\nabla_\mu\nabla_\mu F-\kappa T_{\mu\mu}}{g_{\mu\mu}}
\end{equation}
with no summation on repeated indices, is independent of the index $\mu$. From $K_{[t]}-K_{[r]}=0$, we have:
\begin{equation}
2F\frac{X'}{X}+rF'\frac{X'}{X}-2rF''+2r\kappa\frac{X}{A}(\rho+p)=0
\label{1}
\end{equation}
where $X(r)=A(r)B(r)$ and a prime denotes the derivative with respect to $r$. Also $K_{[t]}-K_{[\theta]}=0$ results in:
\begin{equation}
A''+\left(\frac{F'}{F}-\frac{X'}{2X}\right)\left(A'-\frac{2A}{r}\right)-\frac{2A}{r^2}+\frac{2X}{r^2}+2\kappa\frac{X}{F}(\rho+p)=0
\label{2}
\end{equation}
Furthermore, from the conservation equation of energy momentum tensor, one obtains:
\begin{equation}
\frac{A'}{A}=-\frac{2p'}{\rho+p}
\label{3}
\end{equation}
And also the corresponding curvature scalar is given by:
\begin{equation}
R=-\frac{2}{r^2}+\frac{1}{X}\left[A''+4\frac{A'}{r}-\frac{X'}{X}\left( \frac{A'}{2}+\frac{2A}{r}\right) +\frac{2A}{r^2}\right]
\label{R}
\end{equation}
Equations (\ref{state}), (\ref{1}), (\ref{2}), (\ref{3}) and (\ref{R}) completely determine the unknown quantities $A$, $X$, $p$, $\rho$ and $R$. These are the relativistic equations of stellar structure in $f(R)$ theory of gravity.

From a mathematical point of view, knowing the functional form of $f(R)$, equation (\ref{R}) shows that the functions $F(r)$ and $F'(r)$ depend on $A(r)$ and $B(r)$ in a complicated and non-linear manner. Therefore we can not expect to obtain the analytical solutions of equations (\ref{1}), (\ref{2}) and (\ref{3}) in a closed form. Although one could use the numerical \cite{kain} or approximate methods such as the Newtonian approximation \cite{kain} or perturbation method \cite{cap}.  In the Newtonian limit, assuming a polytropic equation of state of matter, $p\sim\rho^\gamma$, these equations lead to a modified version of Poisson and Lan\'e--Emden equations\cite{Capo}.

Here we are interested in  obtaining some analytical solutions in a different way. In order to do this, we follow Tolman's method \cite{tol} which is developed to provide a variety of explicit solutions of Einstein's equations for a static sphere of fluid. Assuming matter is represented by a perfect fluid, instead of using the equation of the state of matter, Tolman introduces a relation for the metric components or their dependence to radial coordinate, in order to solve the set of equations easily.  Of course, it must be noted that this procedure leads to some expression for the density and the pressure of the perfect fluid which may not be physically acceptable for a star \cite{tol}. 

For $f(R)$ equations, it is necessary to consider two assumptions. One is the equation of the state of the fluid and the other is the form of $f(R)$. Logically these two assumptions should be given to have a completely defined model. We should do this in another way. Two relations for $X$ and $F$ are proposed and then the equations are solved. As a byproduct of the solution, the equation of state and $f(R)$ are obtained. However it has to be noted here that not all solutions lead to physically acceptable equation of  state and also that not for all solutions it is practical to obtain the form of $f(R)$ analytically.

In this paper we discuss two kinds of assumptions: $X=X_0$, $F=F_0r^n$ (class I) and $X=X_0r^m$, $F=F_0r^n$ (class II) where $X_0$, $F_0$, $n$ and $m$ are arbitrary constants. For any choice, a number of interesting solutions for $A$, $p$ and $\rho$ along with the corresponding forms of $f(R)$ are obtained. To specify the integration constants, one must specify the boundary conditions. For a star in equilibrium, it is physical to assume that the pressure drops from its central value to  zero at the surface of star, $R_s$. Moreover the metric needs to be matched with the outside vacuum solution at the surface of the star, according to the Israel junction conditions.  
\subsection{Class I solutions: $X=X_0$, $F=F_0r^n$}
A look at equations (\ref{1})--(\ref{3}), suggests that choosing $X=X_0=\textit{Constant}$ simplifies the equations. Then on choosing $F=F_0r^n$ and combining equations (\ref{1}) and (\ref{2}) we have
\begin{equation}
A''+\frac{nA'}{r}+\frac{2n^2-4n-2}{r^2}A+\frac{2X_0}{r^2}=0
\end{equation}
with the general solution
\begin{equation}
A=A_0+A_1r^m
\end{equation}
where
\begin{equation}
A_0=-\frac{X_0}{n^2-2n-1}
\end{equation}
and
\begin{equation}
m=\frac{1}{2}\left(1-n\pm\sqrt{9+14n-7n^2}\right)
\end{equation}
and $A_1$ is a constant.

Finally equations (\ref{1}) and (\ref{3}) leads to
\begin{equation}
p=p_0+p_1r^{m+n-2}
\end{equation}
where 
\begin{equation}
p_1=-\frac{F_0mn(n-1)A_1}{2\kappa X_0(n+m-2)}
\end{equation}
and $p_0$ is a constant, and also
\begin{equation}
\rho=-p_0+\rho_1r^{n-2}+\rho_2r^{m+n-2}
\end{equation}
where
\begin{equation}
\rho_1=-\frac{F_0n(n-1)}{\kappa (n^2-2n-1)}
\end{equation}
and
\begin{equation}
\rho_2=\frac{F_0n(n-1)(2n+3m-4)A_1}{2\kappa X_0(n+m-2)}
\end{equation}

In addition one can simply see that
\begin{equation}
R=\frac{R_1}{r^2}+R_2r^{m-2}
\end{equation}
with
\begin{equation}
R_1=-\frac{2n(n-2)}{n^2-2n-1}
\end{equation}
and
\begin{equation}
R_2=\frac{(m^2+3m+2)A_1}{X_0}
\end{equation}

Although these are exact solutions of the $f(R)$ equations, but the form of $f(R)$ only can be reconstructed for special choices of $n$. These are listed in Table(\ref{t1}).
\begin{sidewaystable}
\caption{The first class of solutions with $X=X_0$ and $F=F_0r^n$}
\centering
\begin{tabular}{c||ccccccc}
& $A(r)$ & $p(r)$ & $\rho(r)$ & $R(r)$ & $f(R)$ & $v_s^2=\frac{dp}{d\rho}$ & acceptable? \\ 
\hline 
\hline 
I ($n=0$, $m=-1$) & $X_0+\frac{A_1}{r}$ & constant & constant & 0 & any regular & -1 & no \\ 
II ($n=0$, $m=2$) & $X_0+A_1r^2$ & constant & constant & $\frac{12A_1}{X_0}$ & any & -1 & no \\ 
III ($n=1$, $m=-2$) & $\frac{X_0}{2}+\frac{A_1}{r^2}$ & constant & constant & $-\frac{1}{r^2}$ & $f_0-2F_0\sqrt{-R}$ & -1 & no \\ 
IV ($n=1$, $m=2$) & $\frac{X_0}{2}+A_1r^2$ & constant & constant & $\frac{12A_1}{X_0}-\frac{1}{r^2}$ & $f_0-2F_0\sqrt{\frac{12A_1}{X_0}-R}$ & -1 & no \\ 
V ($n=2$, $m=1$) & $\frac{X_0}{2}+A_1r$ & $p_0-\frac{F_0A_1r}{\kappa X_0}$ & $-p_0+\frac{2F_0}{\kappa}+\frac{3F_0A_1r}{\kappa X_0}$ & $\frac{6A_1}{X_0r}$ & $f_0-\frac{36A_1^2F_0}{X_0^2R}$ & $-\frac{1}{3}$ & no \\
VI ($n=2.5$, $m=-1$) & $-4X_0+\frac{A_1}{r}$ & $p_0-\frac{15F_0A_1}{4\kappa X_0\sqrt{r}}$ & $-p_0-\frac{15F_0\sqrt{r}}{\kappa}+\frac{15F_0A_1}{2\kappa X_0\sqrt{r}}$ & $-\frac{10}{r^2}$ & $f_0-\frac{10F_0(-10R)^{-1/4}}{4}$ & $\frac{-A_1}{2A_1+4X_0r}$ & thick shell (see Discussion section) \\ 
\end{tabular} 
\label{t1}
\end{sidewaystable}

Solutions I, II, III, IV, and V are not physically acceptable, because they lead to a negative sound velocity squared $v_s^2=dp/d\rho$. The last solution is acceptable for a thick shell. The minimum and maximum radii of the shell would be derived from the conditions that $0\le v_s^2\le 1$ and $\rho\ge 0$ and $p\ge 0$. As a result we have a thick shell:
\begin{equation}
\frac{3A_1}{A_0}\le r\le \frac{p_1^2}{p_0^2}
\end{equation}
It has to be noted here that to have a physically complete solution, this thick shell solution have to be matched to the vacuum solution at $r=3A_1/A_0$ and $r=p_1^2/p_0^2$. The Israel junction conditions for $f(R)$ theory can be found in the literature\cite{I1,I2,I3}. Choosing the coordinates so that the metric is in the form:
\begin{equation} 
ds^2=N^2dz^2+h_{ij}dx^idx^j
\end{equation}
where $z$ is in the jump direction, the Israel junction conditions are:
\begin{equation}
\Delta h_{ij}=0
\label{j1}
\end{equation}
\begin{equation}
\Delta R=0
\label{j2}
\end{equation}
\begin{equation}
\Delta K=0
\label{j3}
\end{equation}
\begin{equation}
F(R)\Delta \bar{K}_{ij}=\bar{S}_{ij}
\label{j4}
\end{equation}
\begin{equation}
3\frac{dF(R)}{dR}\Delta \partial_z R=N S
\label{j5}
\end{equation}
in which $K_{ij}$ is the extrinsic curvature defined as:
\begin{equation}
K_{ij}=-\frac{1}{2N}\partial_z h_{ij}
\end{equation}
$K$ is its trace, $\bar{K}_{ij}$ is the traceless part of $K_{ij}$, $S_{ij}$ is the energy--momentum tensor on the boundary (the thin shell), $S$ is its trace and $\bar{S}_{ij}$ is the traceless part of $S_{ij}$.
\subsection{Class II solutions: $X=X_0r^m$, $F=F_0r^n$}
Choosing 
\begin{equation}
F=F_0r^n
\end{equation}
and
\begin{equation}
X=X_0r^m
\end{equation}
the equations (\ref{1})--(\ref{3}) can be solved leading to:
\begin{equation}
A=A_0r^m+A_1r^{l_1}+A_2r^{l_2}
\end{equation}
where 
\begin{equation}
A_0=X_0\delta;\ \ \ \ \ \ \ \delta=\frac{-4}{4n^2-8n+m^2-4m-4}
\end{equation}
and $l_1$ and $l_2$ are the roots of the equation:
\begin{equation}
l^2+(n-1-m/2)l+(2n^2-4n-nm-m-2)=0
\end{equation}
and $A_1$ and $A_2$ are two constants. 

Some special choices of $n$ and $m$ allows one to reconstruct the form of $f(R)$ which are listed in Table(\ref{t2}).
\begin{sidewaystable}
\centering
\caption{The second class of solutions with $X=X_0r^m$ and $F=F_0r^n$}
\begin{tabular}{c||ccccccc}
& $A(r)$ & $p(r)$ & $\rho(r)$ & $R(r)$ & $f(R)$ & $v_s^2=\frac{dp}{d\rho}$ & acceptable? \\ 
\hline 
\hline 
I & $A_0r^m$ & $p_0+p_1r^{n-2}$ & $-p_0-\frac{2n+m-4}{m}p_1r^{n-2}$ & $\frac{R_1}{r^2}$ & $f_0+\frac{F_0R_1^{n/2}}{1-n/2}R^{1-n/2}$\ \  for $n\ne 2$ & $\frac{-m}{m+2n-4}$ & some $m,n$ \\ 
&&&$\frac{4\kappa X_0p_1}{F_0A_0}=$&$R_1=$&$f_0+F_0R_1\ln R$\ \  for $n=2$&&\\
&&&$-\frac{m(2n(n-1)-m(n+2))}{n-2}$&$\frac{-4m^2+4m-8n^2+16n}{m^2-4m+4(n^2-2n-1)}$&&&\\
\hline
II & $A_0r^{10}+A_1r^8$ & $p_0+\frac{75F_0}{13\kappa}\frac{1}{r}-20\frac{F_0A_1}{\kappa X_0}\frac{1}{r^3}$ & $-p_0-\frac{60F_0}{13\kappa}\frac{1}{r}+5\frac{F_0A_1}{\kappa X_0}\frac{1}{r^3}$ & $\frac{-314}{13r^2}+\frac{30A_1}{X_0r^4}$ & $f_0-\frac{4(1-aR+\sqrt{1+aR})}{3a}\sqrt{\frac{-1+\sqrt{1+aR}}{R}}$ & $\frac{5X_0r^2-52A_1}{-4X_0r^2+13A_1}$ & thick shell \\ 
&&&&& where $a=30\left(\frac{13}{157}\right)^2\frac{A_1}{X_0}$&& (See Discussion section)\\
\hline
III & $A_0r^{10}+A_1r^8$ & $p_0+\frac{20F_0}{7\kappa}\ln r-\frac{16F_0A_1}{\kappa X_0}\frac{1}{r^2}$ & $-p_0-\frac{4F_0}{7\kappa}-\frac{20F_0}{7\kappa}\ln r+$ & $\frac{-45}{7r^2}+\frac{30A_1}{X_0r^4}$ & $f_0+2\ln (-1+\sqrt{1+aR})+$ & $\frac{X_0r^2+11.2A_1}{-X_0r^2-5.6A_1}$ & thick shell \\ 
&&&$\frac{8F_0A_1}{\kappa X_0}\frac{1}{r^2}$&& $2\sqrt{1+aR}$ where $a=120\left(\frac{13}{157}\right)^2\frac{A_1}{X_0}$&& (See Discussion section) \\
\hline
IV & $A_0r^4+A_1r^2$ & $p_0$ & $-p_0$ & $\frac{-4}{r^2}$ & $f_0-\frac{16F_0}{R}$ & $-1$ & no \\ 
\hline
V & $A_0r^6+A_1r^4$ & $p_0-\frac{9F_0}{10\kappa}r^2+\frac{12F_0A_1}{\kappa X_0}\ln r$ & $-p_0-\frac{6F_0A_1}{\kappa X_0}+$ & $\frac{-23}{5r^2}+\frac{6A_1}{X_0r^4}$ & $f_0-\frac{aF_0}{2R}+\frac{bF_0}{4}\ln R\pm$ & $\frac{-3X_0r^2+20A_1}{5X_0r^2-20A_1}$ & thick shell \\ 
&&&$\frac{3F_0}{2\kappa}r^2-\frac{12F_0A_1}{\kappa X_0}\ln r$&& $\frac{23\sqrt{a+bR}}{10R}\pm\frac{b}{2}\tanh^{-1}\sqrt{1+bR/a}$ && (See Discussion section) \\
&&&&& where $a=\left(\frac{23}{5}\right)^2$ and $b=\frac{24A_1}{X_0}$ &&\\
\end{tabular} 
\label{t2}
\end{sidewaystable}
Solution IV gives a negative sound velocity squared and thus is not acceptable. Solutions II, III, and V are acceptable for a thick shell.
Again Israel junction conditions should be applied on it to have a complete solution. Finally solution I, with some values of $n$ and $m$ gives a physical acceptable solution for a star.
\section{Matching with exterior solution}
A look at Tables (\ref{t1}) and (\ref{t2}), shows that only the solution I of Table (\ref{t2}) is appropriate for a star. To have a physical sphere of fluid surrounded by empty space, we set $n<2$, so that the energy density and pressure decrease in a monotonic way with decreasing $r$, and the equation of state can be written as:
\begin{equation}
p=\omega\rho+p_0(\omega+1)
\end{equation}
where $\omega=\frac{m}{2n+m-4}$.
Moreover in spite of obtaining infinite values for central density and pressure at $r=0$, they are  positive definite provided that $p_0<0$, $p_1>0$ and $\omega>0$. The pressure vanishes at the surface of the star $r=R_s$, and thus $R_s=(\frac{-p_0}{p_1})^{1/(n-2)}$. We have to have subluminal speed of sound and thus $0<\omega<1$ and thus $0<m<2-n$.

Solution I specialize us  to power law lagrangian plus a cosmological constant. Since $f(R)$ models are of interest because of the fact that they can explain the late--time acceleration of the universe without the need for a cosmological constant, we set $f_0=0$. Fortunately, for this case in 
the absence of cosmological constant, there are two spherically symmetric vacuum static solutions in the literature. One is  proposed in \cite{clif1} and another is given by \cite{vil6}. Therefore one can consider the matching of the interior metric to each of these forms of the exterior solutions at the surface of the sphere. Here we take the first choice. 

The exterior solution is thus\cite{clif1}:
\begin{equation}
A_{ext}=r^\alpha +C r^\beta
\label{e1}
\end{equation}
\begin{equation}
X_{ext}=\gamma r^\alpha
\label{e2}
\end{equation}
where $C$ is a constant, $\alpha=\frac{2n(n-1)}{n+2}$, $\beta=\frac{-4n-2}{n+2}$ and $\gamma=1-\frac{n(2n^3-2n^2-5n-4)}{(n+2)^2}$. The parameter $n$ is subject to some constraints. The existence of chameleon--like mechanism (named by \cite{Rosh} so, to denote that the chameleon behavior in the Jordan frame differs slightly from the Einstein frame.), light deflection experiment and the perihelion precession of Mercury leads to the following bound on $n$ \cite{clif1, clif2,Rosh, clifton}:
\begin{equation}
-1.6954\times 10^{-17}\le n \le 6.7800 \times 10^{-17}
\label{b}
\end{equation}

Using the exterior solution (\ref{e1}) and (\ref{e2}) and the interior solution I of table (\ref{t2}), the junction conditions given by the equations  (\ref{j1})--(\ref{j3}), would be simplified into:
\begin{equation}
\Delta h_{ij}=0 \Longrightarrow A_0R_s^m=R_s^\alpha+CR_s^\beta
\end{equation}
\begin{equation}
\Delta K=0 \Longrightarrow (m+4)\sqrt{\frac{\gamma\delta}{A_0}}=R_s^{(m-\alpha)/2}\left[ 4+\frac{\alpha R_s^{\alpha-m}+C\beta R_s^{\beta-m}}{A_0}\right]
\end{equation}
\begin{equation}
\Delta R=0 \Longrightarrow \gamma\delta(m+2)^2=(\alpha+2)^2
\end{equation}
The first two equations determine the constants $A_0$ and $C$ in terms of $R_s$, $n$ and $m$, while the third equation can be used to find $m$ in terms of $n$. It can be solved exactly and there are two solutions:
\begin{equation}
m_1=\frac{2n(n-1)}{n+2}
\end{equation}
and
\begin{equation}
m_2=\frac{-2(n^2-3n+2)}{n-4}
\end{equation}
It has to be noted that only the ratio $\frac{p_1}{F_0}$ is fixed according to the relation given in table (\ref{t2}). Therefore given a specific form of $f(R)$ which fixes $F_0$, one gets the constant $p_1$.
For $m_1$ one can simply see that $p_1=0$ and this wipes out the star solution. For $m_2$ we find:
\begin{equation}
p_1=-\frac{4F_0}{\kappa}\frac{(n-1)^3(n-4)^4}{(n-4)^2(n^2-2n-1)+(n-1)(n^2-4)(n-3)}
\end{equation}
Also $p_0$ can be derived from the definition of $R_s$ as:
\begin{equation}
p_0=-p_1R_s^{n-2}
\end{equation}

\begin{figure}[htb]
\begin{center}
\epsffile{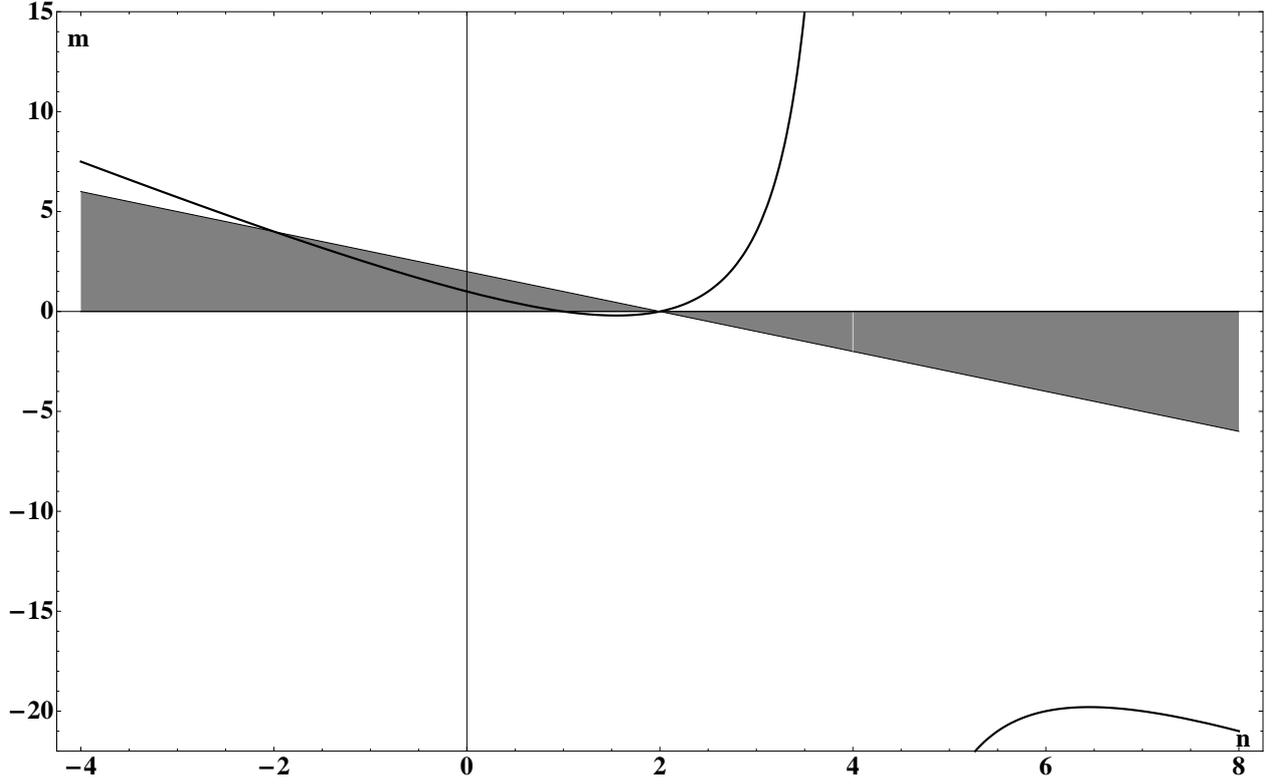}
\end{center}
\caption{$m$ as a function of $n$.  Those values of $m$ and $n$ that lies in the shaded region lead to a subluminal sound velocity.}
\label{f1}
\end{figure}

As stated before, to have subluminal sound velocity we have to have $0<m<2-n$. The acceptable values of $n$ and $m$ are plotted in Figure (\ref{f1}). Also taking into account the bound given by equation (\ref{b}) we see that  values of $n$  are very small and thus:
\begin{equation}
 m \simeq 1-\frac{5}{4}n
\end{equation}

The last two junction conditions (\ref{j4}) and (\ref{j5}) determine the surface energy--momentum tensor. Equation (\ref{j5}) gives:
\begin{equation}
S=0
\end{equation}
while equation (\ref{j4}) simplifies to:
\begin{equation}
\bar{S}_{00}=A_0F_0R_s^n\left(\frac{m\sqrt{\delta}}{2}R_s^{m-1}+2\sqrt{\frac{A_0}{\gamma}}R_s^{(m-\alpha)/2}-\frac{m+4}{2}\sqrt{\delta}\right)
\end{equation}
\begin{equation}
\bar{S}_{22}=F_0R_s^{n+1}\left(\sqrt{\delta}-\sqrt{\frac{A_0}{\gamma}}R_s^{(m-\alpha)/2}\right)
\end{equation}
\begin{equation}
\bar{S}_{33}=\bar{S}_{22} \sin ^2\theta
\end{equation}
$\bar{S}_{00}$ is the energy density and $\bar{S}_{22}$ and $\bar{S}_{33}$ are the tension densities on the star surface.
\epsfxsize=6.5in
\section{Discussion}  
In this paper we found some static spherically symmetric interior solutions of $f(R)$ theory of gravity. Of the eleven solutions, listed in table (\ref{t1}) and (\ref{t2}), only one solution is physically acceptable for a star. 
In obtaining the solutions, we restricted ourselves to the solutions which have two properties. First, they are in the closed form and are obtained without any numerical or approximation method for solving the coupled high order differential equations of $f(R)$ theory. Second, the corresponding form of $f(R)$ function can be reconstructed. This means that we have not reported those solutions for which there are algebraic
relations for $R(r)$ and $f(r)$, without the possibility for solving them to get $f(R)$. 

Special attention is paid to the star solution, and it is matched to existing exterior solution and thus as a result one has a complete solution for a star and its exterior region for power law $f(R)$ theories. The junction conditions are applied on this star solution and the acceptable values of the constants are derived.

It has to be noted that we are not able to apply the junction conditions for our thick shell solutions. This is because of the fact that we do not know the exterior solution for such $f(R)$'s. Moreover those forms of $f(R)$ are not of physical interest.

\textbf{Acknowledgement} This work is partly supported by a grant from university of Tehran. One of us (F. Shojai) wishes to thanks the center of excellence of department of physics on the structure of matter for supporting this work.

\end{document}